\documentclass[lettersize,journal]{IEEEtran}
\usepackage{amsmath,amsfonts}
\usepackage{algorithmic}
\usepackage{algorithm}
\usepackage{array}
\usepackage[caption=false,font=normalsize,labelfont=sf,textfont=sf]{subfig}
\usepackage{textcomp}
\usepackage{stfloats}
\usepackage{url}
\usepackage{verbatim}
\usepackage{graphicx}

\usepackage{siunitx}

\usepackage{hyperref}

\usepackage[backend=biber,style=ieee,dashed=false, doi=false, mincitenames=1, maxcitenames=1]{biblatex}
\AtBeginBibliography{\footnotesize}

\usepackage{lmodern}

\usepackage{layouts}

\usepackage{tikz}

\PassOptionsToPackage{dvipsnames}{xcolor}
\usepackage[]{xcolor}
\usepackage{tikz}
\usetikzlibrary{backgrounds}
\usetikzlibrary{arrows,shapes}
\usetikzlibrary{tikzmark}
\usetikzlibrary{calc}

\usepackage{pgfplots}
\DeclareUnicodeCharacter{2212}{−}
\usepgfplotslibrary{groupplots,dateplot}
\usetikzlibrary{patterns,shapes.arrows}
\pgfplotsset{compat=newest}
\usepackage{standalone}

\usepackage{amsmath}
\usepackage{amsthm}
\usepackage{amssymb}
\usepackage{mathtools, nccmath}
\usepackage{wrapfig}
\usepackage{comment}

\usepackage{blindtext}

\usepackage[]{adjustbox}
\usepackage{standalone}

\usepackage{graphicx}
\usepackage{xspace}

\usepackage{array}
\usepackage{ragged2e}
\newcolumntype{P}[1]{>{\RaggedRight\hspace{0pt}}p{#1}}
\newcolumntype{X}[1]{>{\RaggedRight\hspace*{0pt}}p{#1}}

\usepackage{tcolorbox}

\usepackage{booktabs}
\usepackage{graphicx}
\usepackage{lscape}

\usepackage{tikz}
\usetikzlibrary{arrows,shapes,positioning,shadows,trees,mindmap}
\usepackage[edges]{forest}
\usetikzlibrary{arrows.meta}
\colorlet{linecol}{black!75}
\usepackage{xkcdcolors} % xkcd colors

\usepackage{tikz}
\usetikzlibrary{backgrounds}
\usetikzlibrary{arrows,shapes}
\usetikzlibrary{tikzmark}
\usetikzlibrary{calc}
\newcommand{\highlight}[2]{\colorbox{#1!17}{$\displaystyle #2$}}
\usepackage{bm}

\renewcommand{\highlight}[2]{\colorbox{#1!17}{#2}}

\usepackage{orcidlink}

\usepackage{glossaries} % Glossaries
\makeglossaries

\begin{document}

% Chanhe space between lines
% \linespread{0.8}

\newacronym{rbs}{RBS}{Reference Broadcast Synchronization}
\newacronym{ftsp}{FTSP}{Flooding Time Synchronization Protocol}
\newacronym{tspn}{TSPN}{Time Synchronization Protocol for Sensor Networks}
\newacronym{ftsp}{FTSP}{Flooding Time Synchronization Protocol}
\newacronym{rtc}{RTC}{Real-Time Clock}
\newacronym[plural=IMUs]{imu}{IMU}{Inertial Measurement Unit}
\newacronym{ble}{BLE}{Bluetooth Low Energy}
\newacronym{soc}{SoC}{System on Chip}
\newacronym{fpu}{FPU}{Floating Point Unit}
\newacronym{ppi}{PPI}{Programmable peripheral interconnect}
\newacronym{cpu}{CPU}{Central Processing Unit}
\newacronym[plural=WBANs, firstplural= Wireless Body Area Networks (WBANs)]{wban}{WBAN}{Wireless Body Area Network}
\newacronym{pll}{PLL}{Phase Locked Loop}
\newacronym{dcu}{DCU}{Data Capturing Unit}
\newacronym{dof}{DoF}{Degrees of Freedom}
\newacronym{dmp}{DMP}{Digital Motion Processor}
\newacronym{fifo}{FIFO}{First In First Out}
\newacronym{wsn}{WSN}{Wireless Sensor Network}
\newacronym{emg}{EMG}{Electromyography}
\newacronym{gpio}{GPIO}{General Purpose Input/Output}
\newacronym{ntp}{NTP}{Network Time Protocol}
\newacronym{mac}{MAC}{Medium Access Control}
\newacronym{tpsn}{TPSN}{Timing-sync Protocol for Sensor Networks}
\newacronym{ram}{RAM}{Random Access Memory}
\newacronym{lipo}{LiPo}{Lithium polymer}
\newacronym{adc}{ADC}{Analog to Digital Converter}
\newacronym{mcu}{MCU}{microcontroller unit}

\title{Low-Power Synchronization for Multi-IMU WSNs}

\author{IEEE Publication Technology,~\IEEEmembership{Staff,~IEEE,}
        % <-this % stops a space

\author{
    \IEEEauthorblockN{Jona Cappelle \orcidlink{0000-0002-2084-9875}, Sarah Goossens \orcidlink{0000-0001-7525-1179}, Lieven De Strycker \orcidlink{0000-0001-8172-9650}, Liesbet Van der Perre \orcidlink{0000-0002-9158-9628}\\}
    \IEEEauthorblockA{
        \textit{KU Leuven, ESAT-WaveCore, Ghent Technology Campus}   \\
                        B-9000 Ghent, Belgium                                                               \\
                        name.surname@kuleuven.be                                                        }
}

\thanks{This paper was produced by the IEEE Publication Technology Group. They are in Piscataway, NJ.}% <-this % stops a space
\thanks{Manuscript received April 19, 2021; revised August 16, 2021.}}

\markboth{Journal of \LaTeX\ Class Files,~Vol.~14, No.~8, August~2021}%
{Shell \MakeLowercase{\textit{et al.}}: A Sample Article Using IEEEtran.cls for IEEE Journals}


\maketitle

\begin{abstract}
% Intro
Wireless time synchronization is one of the most important services in a \acrfull{wsn}. \Acrfullpl{imu} are often used in these \acrshortpl{wsn} in healthcare-related treatments. %, which have recently gained a lot of attention.
% What have we made?
We present a low-power, wirelessly synchronized multi-\acrshort{imu} platform. The proposed approach synchronously captures packets from different \acrshortpl{imu} and transmits the data over \acrfull{ble} to a central \acrfull{dcu}.
% What's the goal of this solution?
The contribution of this work is, rather than focussing on the highest possible accuracy, to provide a low-power accurate enough solution for use in a multi-\acrshort{imu} \acrshort{wsn}. 
% What are our contributions?
We examine key factors affecting synchronization accuracy and elaborate on the implementation challenges.
% What's the resulting accuracy?
An accuracy of sub 1 $\mathrm{\mu}$s can be achieved with the approach using 74.8~J$\mathrm{\mathbf{h^{-1}}}$ of energy, while a power-optimized implementation is presented with an accuracy of 200~$\mathrm{\mu}$s and an energy consumption of only 198~mJ$\mathrm{\mathbf{h^{-1}}}$. 
This approach suits the required accuracy and low-power requirements for a multi-IMU system.
% 
%
% \jona{abstract aanpassen, moet duidelijk blijken dat het gaat over een multi sensor solution, met een laag genoeg vermogenverbruik en degelijk genoege synchronizatie accuracy. We streven hier niet naar de hoogst mogelijk haalbare accuracy maar een low genoege power implementatie in een multi-imu systeem. We bespreken de belangrijke parameters en implementatie challenges.}
\end{abstract}


%%%%%%%%%%%%%%%%%%%%%%%%%%%%%%%%%%%%%%%%%%%%%%%%%%%%%%%%%%%%%%%%%%%%%%

\begin{figure*}[hbt!]
    \centering
    \includegraphics[width=\textwidth]{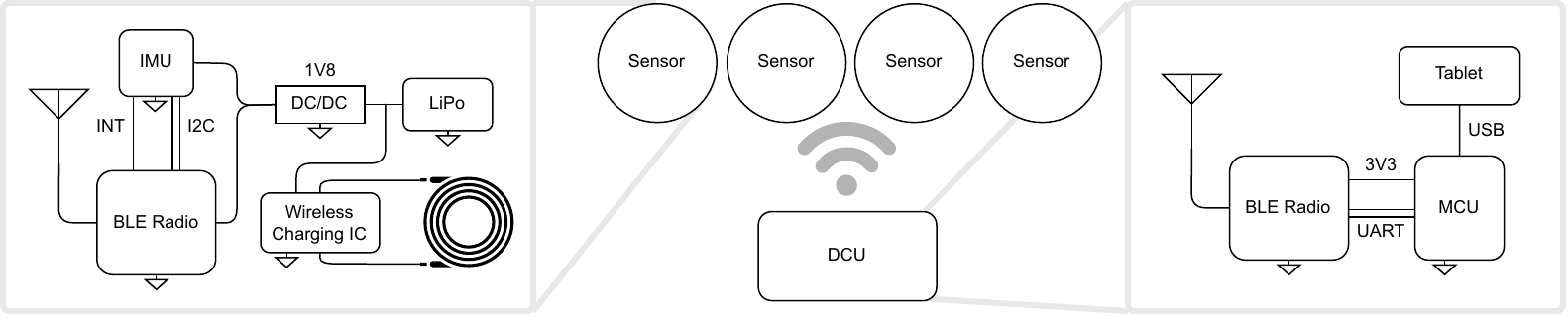}
    \caption{System overview. A central \acrfull{dcu} with multiple wireless \acrfullpl{imu} sensor modules. In the center: top-down view of system. On the left: block diagram of the sensor node. On the right: simplified block diagram of the \acrshort{dcu}.}
    \label{fig:system_overview}
    \vspace{-12pt}
\end{figure*}

% Introduction
% \section{Introduction and State of the art}
% \jona{Overwegen om intro en state of the art samen te brengen}
% \jona{Verwijzing naar github}
% \jona{deel van problem statement (sectie 4) naar introductie}

\IEEEPARstart{W}{ireless} synchronization is becoming an essential service in \glspl{wsn}. In particular, in \glspl{wban}, an increasing amount of sensors is being used to conduct simultaneous synchronized measurements on different parts of the body. %\liesbet{open question: position the contribution for WBAN and then clarify it could be applied in other contexts, or as is done now claim the generic approach and focus on the WBAN application as a case?}
For physiotherapists, it can be beneficial to use \glspl{imu}, instead of high-tech and expensive equipment, to study the long-term evolution of certain symptoms. 
One wants samples from all \gls{imu} to be synchronized, enabling the direct comparison of samples at various locations on the body.
Essentially all measured body movements are contained within frequency components below \SI{20}{Hz}~\cite{imu_what_frequency}.
% For position estimation, higher sampling rates between \SI{200}{Hz} - \SI{300}{Hz} offer a significant advantage~\cite{foot_tracking}.
% Since the \glspl{imu} will be used for movement analysis, according to Nyquist theorem a sampling frequency of minimal \SI{40}{Hz} is sufficient. 
According to the Nyquist theorem, a sampling frequency of at least \SI{40}{Hz} should be sufficient. 
We consider the \gls{wban} case requiring synchronized sampling and design a system with a sampling frequency of \SI{100}{Hz}. 
% We dimensioned our system around a sampling frequency of \SI{100}{Hz}. 
To guarantee packet-level synchronization accuracy, a maximum error of half a sampling period, i.e., \SI{5}{ms}, is admitted.

% \subsection{Existing Standards for Wireless Synchronization}
Existing solutions for wireless synchronization have been proposed.
% Three mainly used synchronization methods are:
% \gls{ntp},
% \gls{rbs},
% \gls{tspn}, and the
% \gls{ftsp}~\cite{ntp, rbs, tspn, flooding_timesync}.
%
%
\Gls{ntp}~\cite{ntp} is broadly known for keeping a synchronized time on the internet, and can guarantee an accuracy within a few milliseconds.
In \gls{rbs}~\cite{rbs} the master sends broadcast reference messages, the peripheral informs in turn other peripherals about time change. 
% The reference broadcast does not contain a timestamp, instead, 
Receivers here use the arrival time as a means to adjust their local clocks. Compared to \gls{ntp}, \gls{rbs} eliminates the non-deterministic delays associated with the master, resulting in microsecond level time accuracy.
In \gls{tpsn}~\cite{tspn}, synchronization is based on a handshake, using 2-way communication. Timestamp generation is done at the \gls{mac} layer. Therefore, it achieves an accuracy twice that of \gls{rbs}.
The \Gls{ftsp}~\cite{flooding_timesync} reduces communication overhead by synchronizing nodes based on a single radio message. To further reduce the synchronization error, \gls{ftsp} adds drift compensation, combined with \gls{mac}-layer timestamping. Accuracies of around \SI{1}{\micro\second} are feasible.
%
% A problem that may arise here, is the enormous amount of beacon packets being transmitted to keep the synchronization error small.%, in a way that may hinder the normal wireless communication, also known as the broadcast storm problem~\cite{broadcast_storm_problem}.

% \subsection{Specific \gls{ble} Implementations}
%
In \gls{ble}, 
% Typical synchronization approaches, such as using GPS~\cite{gnss_fsync}, however resulting in very accurate synchronization, are not possible in a small and low-power system.
%
using the connection-based event to synchronize multiple wireless sensor nodes is not very accurate. Since an application-level event is used to trigger a timestamp capture, an accuracy of $\pm$~\SI{750}{\micro\second} is achieved~\cite{ble_sync_connection_based_event}.
% In \cite{ble_sync_connection_based_event} authors explore the possibility of using the connection-based event to synchronize multiple wireless sensor nodes. They use an application level event at the time of \gls{ble} connection establishment to trigger a timestamp. They show an accuracy of $\pm$ \SI{750}{\micro\second}.
In \cite{multi-unit-wireless-platform}, a combination of \gls{rbs} and \gls{ftsp} is used. The timestamp generation is based on a \gls{rtc} timer. With no attention given to hardware timestamping, an accuracy of \SI{30}{\micro\second} was achieved.
% presented a synchronized multi-unit platform for long term activity monitoring with a maximal synchronization error of \SI{30}{\micro\second}. The authors combine an \gls{emg} and \gls{imu} sensors and synchronize them wirelessly using a simplified version of a combination of \gls{rbs} and \gls{ftsp}. The timestamp generation is based on a \gls{rtc} timer. No attention has been given to hardware timestamping the received packet at the peripheral.
% In \cite{gnss_fsync}, they propose using a GPS for an accurate clock and using the FSYNC input of the \gls{imu} to add timestamping to \gls{imu} samples.
Authors in~\cite{cheepsync} use broadcasting packets to transmit the current timestamp value and the aggregate delay incurred during the transmission of the previous packet. A minimum of 2 broadcast packets needs to be sent to synchronize clocks. While also applying clock drift correction, an accuracy of \SI{10}{\micro\second} was achieved.
In~\cite{sync_current_consumption} specific current consumption patterns are used as a means of hardware timestamping. Results show an accuracy of \SI{17}{\micro\second}, however, requiring additional hardware components.
BlueSync is presented in~\cite{bluesync}, featuring average errors as low as \SI{330}{ns} per \SI{60}{sec} (single hop).
% The authors present techniques for low-power operation (such as duty-cycling clocks), accurate timestamp generation, etc.
This system, while providing very accurate synchronization, is relatively complex.
%
% In~\cite{bluesync} the authors present BlueSync, the most accurate time synchronization reported to date. BlueSync can achieve average synchronization errors as low as \SI{330}{ns} per \SI{60}{sec} (single hop). They present techniques for low-power operation (such as duty-cycling clocks), accurate timestamp generation etc. This system, however providing very accurate synchronization, is relatively complex.
In conclusion, solutions have been proposed that achieve the required accuracy, yet are not specifically fit for \gls{ble} or require a complex integration. 
In a multi-\gls{imu} system, a lot of data will be transmitted anyway. The added complexity of a more advanced synchronization method does not outweigh the limited overhead of sending an extra sync packet. 
Complementary, e.g., \gls{ntp} and the connection-based event have been proposed, which are much less accurate.

% \liesbet{I think here we need to draw a conlusion: }

% \jona{Expliciet contributies toelichten}
% \jona{er bestaan oplossingen die een bepaalde accuracy halen, dit is de accuracy die we nodig hebben, dat beantwoord deze paper}

% Contributions
% The contributions of this paper are twofold.
We present a multi-unit wireless \gls{imu} \gls{wban} synchronization approach and platform, %\liesbet{a multi- unit ... synchronization approach and platform?} 
and elaborate on synchronization error localization and implementation challenges on off-the-shelf wireless \glspl{soc}.
The synchronization approach presented suits the accuracy and low-power demands required in \gls{wban}. %for a multi-\gls{imu} \gls{wsn} \liesbet{required in WBAN?}. 
% \liesbet{add here 'The embedded software and hardware are shared...?}
The embedded software and hardware are shared through GitHub~\cite{full_code_github}. %\liesbet{zou dit al eerder aankondigen als een bijdrage, zie suggestie boven}
%
% Overview paper
The remainder of the paper is structured as follows. First, an overview of the system is given. The synchronization analysis and methodology are elaborated on in Section~\ref{sec:synchronization} and \ref{sec:implementation_on_ble_soc}. The results are presented in Section~\ref{sec:results}, with Section~\ref{sec:conclusion} concluding this work.

% \section{State of the art}\label{sec:sota}

% In literature, low-power and accurate synchronization algorithms are already a widely researched topic. First, existing techniques for wireless synchronization will be discussed, followed by concrete implementations in \gls{ble} systems.

% \newpage
\section{System overview}\label{sec:system-overview}

% \jona{for detailed technical overview, see github}

\begin{figure}[bt]
    \centering
    \includegraphics[width=0.8\columnwidth]{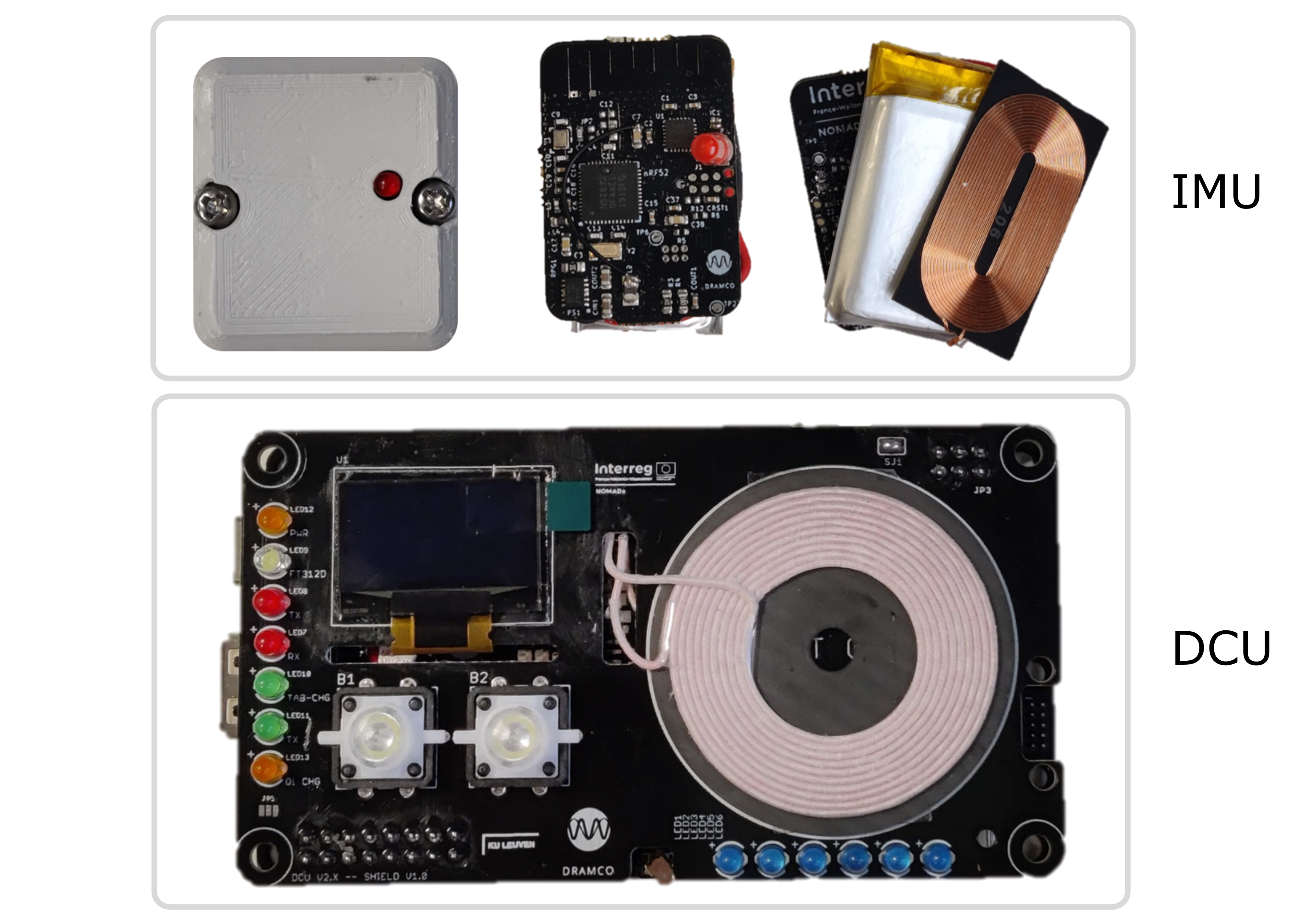}
    \caption{Realized hardware: \Gls{imu} sensor node with \gls{ble} connectivity, powered by a small \gls{lipo} battery, and wirelessly rechargeable (Top). Battery powered DCU for tablet communication, including a \gls{ble} radio, an \gls{mcu} for processing, and Qi wireless charger (bottom).}
    \label{fig:dcu_and_sensor}
    \vspace{-12pt}
\end{figure}

The considered \gls{wsn} system consists of a central \gls{dcu} (master) and a number of \gls{imu} sensors (peripherals), operating in a star network topology. The \gls{dcu} hosts both a \textit{Nordic nRF52832} \gls{ble} \gls{soc} and a \textit{STM32H743ZIT6} \gls{mcu}, which in turn communicates with a tablet and is used for further data processing.
The peripherals consist of \gls{imu} sensor modules with the same \gls{ble} \gls{soc}, which features an ARM Cortex M4+ processor with \gls{fpu} running at \SI{64}{MHz}. It includes a radio module for wireless communications at \SI{2.4}{GHz} and supports \gls{ble}, NFC, ANT and proprietary protocols. The \textit{TDK InvenSense ICM20948}~\cite{icm20948}, despite not having an external clock input, is chosen for its lower power consumption of only \SI{2.5}{mW}.
The \gls{dcu} processes the received \gls{imu} packets and acts as the synchronization master. Up to 8 sensor nodes (peripherals) can be wirelessly connected and synchronized. \Gls{ble} is chosen as the most suited communication technology for its rather high data throughput while still maintaining a low power consumption.
Fig.~\ref{fig:system_overview} gives an overview of the system, Fig.~\ref{fig:dcu_and_sensor} depicts the hardware implementation.
%
%
%
% We selected a 9 \gls{dof} \gls{imu} sensor, since the compass, in addition to the gyroscope and accelerometer, can correct for longer-term drift around the yaw axis.
% The \textit{Bosh BNO055}~\cite{BNO055} sensor was considered for its external clock input capabilities. 
% Synchronizing the \gls{imu} clocks could be done by a shared synchronized clock derived from the central \gls{soc}. %The relatively high power consumption of \SI{36.9}{mW} however, made the \textit{BNO055} harder to justify in a low-power sensor node.
 %, due to the use of an internal FPGA-based \gls{dmp}.% instead of a separate \gls{mcu}. %It features an external FSYNC pin, which could be used for synchronization purposes. \sarah{Maar wordt hier dan niet gebruikt denk ik?}

\section{Synchronization Analysis} \label{sec:synchronization}

% In this system, a number of factors can attribute to poor synchronization accuracy. A subdivision is made into time dependent and time independent errors.

% By using the timeslot API, we can transmit and listen to specific \gls{ble} channels and reserve timeslots for data transmission.

% Timestamping accuracy: How are the timestamps recorded?
% - On application level: capture when the radio generates an interrupt, here we experience several delays (software delays, processor speed) and time sync will thus be less accurate
% - We do this based on \gls{ppi} when the timeslot starts: \gls{ppi} enables autonomous interconnect with predefined behaviour between peripherals by using tasks and events, all independent of the \gls{cpu}. By this way, \gls{cpu} frequency and software delays will not have an effect on the synchronization accuracy: a task can trigger an event without using the \gls{cpu}. In~\cite{bluesync} it is shown that such a timestamping method can achieve an accuracy of on average $<$ \SI{50}{ns} with a standard deviation $<$ \SI{40}{ns}, while software interrupt latency is up to 5 timer larger.

% \subsection{Desired accuracy}

% \jona{specificaties van wat we willen bereiken naar boven}

% \subsection{Time Dependent Errors}

% In Fig.~\ref{fig:imu_mismatch_illustration}, 2
Working with low-cost components, one may expect the accuracy and stability of their clocks not to be high. Low cost components do not always have the necessary inputs, i.e. a clock input, to accommodate optimal synchronization.
We performed the following experiment to illustrate the impact and consequent need to perform a dedicated synchronization. When observing two identical \glspl{imu} sampling at \SI{100}{Hz} during 30 minutes and measuring their sampling rate by the interrupt pin (set to toggle when a sample is available),
% Two \glspl{imu} are set to sample at \SI{100}{Hz} during 30 minutes. The sampling frequency of both \glspl{imu} are measured by their interrupt pin, which is set to toggle when a sample is ready. Measurements are conducted using a logic analyzer at \SI{24}{Msamples/s}.
two problems can be identified:
A \textbf{frequency mismatch}: \SI{103.5}{Hz} and \SI{104.4}{Hz} instead of the expected \SI{100}{Hz}. 
    % Since the chosen \gls{imu} has no external crystal input, we have to rely on the internal \gls{pll} for the clock generation and can't correct for this mismatch. After \SI{100}{samples}, the \glspl{imu} drifted almost \SI{10}{ms} from each other. This type of error is also presented in Fig.~\ref{fig:causes_of_error-scematic}. 
A \textbf{frequency instability/drift}: a clear spread on the clock frequency of $\pm$ \SI{2}{Hz}.
%
%
% \begin{figure}[t!]
%     \centering
%     \resizebox{0.9\columnwidth}{!}{
%         \input{imu_internal_clock_freq}
%     }
%     \caption{\Gls{imu} internal clock frequency mismatch. A clear difference in sampling frequency can be observed due to \gls{imu} clock frequency mismatch. \Gls{imu} 1 samples at \SI{103.5}{Hz}, \gls{imu} 2 samples at \SI{104.4}{Hz}.}
%     \label{fig:imu_mismatch_illustration}
% \end{figure}
%
We propose a synchronization approach to address %The proposed synchronization approach addressed\liesbet{als we hier onze oplossing bedoelen zou ik dit actiever formuleren: we propose a synchronization approach to address ...} 
the two problems by a combination of measures: 

% \begin{enumerate}
1) \textbf{Oversampling the \gls{imu}} at 225 Hz.
With each \gls{imu} interrupt, the \gls{fifo} buffer is read, and the last available value is saved on the microcontroller in \gls{ram}. Only the last stored sample is read at a frequency determined by the user. By doing so, a small error (max. \SI{4.4}{ms}) %\liesbet{frequency niet geïntroduceerd, oversampling by a factor that results in ..., in this case we have chosen ...?}
is introduced between actual sampling time and buffered data, which is quasi-random but does not increase over time. 
% This error is also shown in Fig.~\ref{fig:imu_mismatch}.
% the \gls{imu} and buffering the latest samples. 

2) \textbf{Triggering packet transmission using a synchronized clock.}
Compared to using the free-running \gls{imu} clock to trigger packet transmission, we trigger packet transmission based on an accurate synchronized clock (depicted in Fig.~\ref{fig:rtc_sync}), generated by the \gls{ble} \gls{soc}.
% \end{enumerate}

%
% The maximal sampling error at a frequency of \SI{225}{Hz} (max. sampling frequency) is equal to 1 period, i.e. \SI{4.44}{ms}.

% \begin{figure}[b!]
%     \centering
%     \input{timesync_timedomain}
%     \caption{Illustration of varying \gls{imu} sampling mismatch}
%     \label{fig:imu_mismatch}
% \end{figure}

% Equation~\ref{eq:imu_error_225hz} shows the maximal error at a sampling frequency of \SI{225}{Hz}. 
% % The FSYNC pin of the \gls{imu} can not be used to improve synchronization, since it only enables intra-packet and no inter-packet time stamping.

% \begin{align}
%     T_{sampling} &= \dfrac{1}{f_{sampling}} = \dfrac{1}{\SI{225}{Hz}} = \SI{4.444}{ms}\label{eq:imu_error_225hz}
% \end{align}

\begin{figure}[hbt!]
% \centering
    % \vspace{\baselineskip}
% \begin{minipage}{0.5\columnwidth}

\definecolor{aqua}{HTML}{a7f5f2}
\definecolor{color0}{HTML}{9ac7f6}
\definecolor{ashgrey}{rgb}{0.5, 0.5, 0.5}

\begin{equation*}
\hspace{100pt minus 1fil}
\tau = 
\color{purple}
\overbrace{ 
    \tikzmarknode{ts}{\highlight{teal}{ \color{black}  $t_s$ }} \color{black} + \tikzmarknode{ta}{\highlight{orange}{ \color{black}  $t_a$}} \color{black} + \tikzmarknode{ttx}{\highlight{blue}{ \color{black} $t_{tx}$}}  
    \color{black} +  
    \tikzmarknode{tp}{\highlight{purple}{ \color{black}   $t_p$ }} + 
    \tikzmarknode{tr}{\highlight{ashgrey}{ \color{black}   $t_r$ }} + 
    \tikzmarknode{timu}{\highlight{green}{ \color{black}   $t_{IMU}$ }}
    \color{black}
}^{\substack{\text{\sf \footnotesize \textcolor{purple!85}{Total time 
	}} } }
\hfilneg
\end{equation*}

\vspace*{0.5\baselineskip}

\begin{tikzpicture}[overlay,remember picture,>=stealth,nodes={align=left,inner ysep=1pt},<-]

% For "t_{j+1}"
\path (ts.north) ++ (-2,-2em) node[anchor=north west,color=black] (tj1text){\textsf{\footnotesize Send time}};
\draw [color=black!75](ts.south) |- ([xshift=-0.3ex,color=black]tj1text.south west);

\path (ta.north) ++ (-2.5,-3.2em) node[anchor=north west,color=black] (tj1text){\textsf{\footnotesize Access time}};
\draw [color=black!75](ta.south) |- ([xshift=-0.3ex,color=black]tj1text.south west);

% For "t_{j}"
\path (ttx.north) ++ (-3,-4.4em) node[anchor=north west,color=black] (tjtext){\textsf{\footnotesize Transmission time}};
\draw [color=black!75](ttx.south) |- ([xshift=-0.3ex,color=black]tjtext.south west);

\path (tp.north) ++ (-3,-5.6em) node[anchor=north west,color=black] (tjtext){\textsf{\footnotesize Propagation time}};
\draw [color=black!75](tp.south) |- ([xshift=-0.3ex,color=black]tjtext.south west);

\path (tr.north) ++ (-3,-6.8em) node[anchor=north west,color=black] (tjtext){\textsf{\footnotesize Reception time}};
\draw [color=black!75](tr.south) |- ([xshift=-0.3ex,color=black]tjtext.south west);

\path (timu.north) ++ (-3,-8em) node[anchor=north west,color=black] (tjtext){\textsf{\footnotesize IMU sampling time}};
\draw [color=black!75](timu.south) |- ([xshift=-0.3ex,color=black]tjtext.south west);

\end{tikzpicture}

\vspace*{3.75\baselineskip}
% \vspace{\baselineskip}

\caption{Causes of error in time synchronization}
\label{fig:causes_of_error-scematic}
% \vspace{-12pt}
\end{figure}
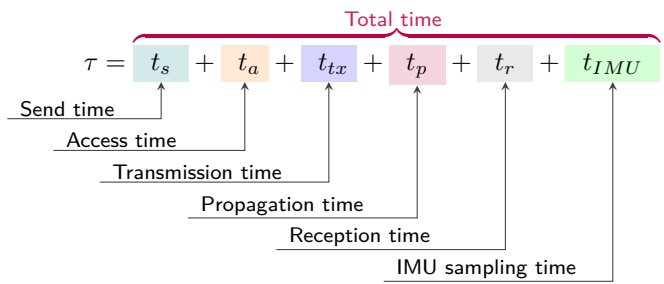

% The crystals on the \textit{nRF} will show similar errors. Unlike the \gls{imu} clock, we can control and synchronize the clocks of the \textit{nRF} over \gls{ble}.
%
% The crystals on the \textit{nRF52832} will show similar errors. Unlike the \gls{imu} clock, we can control and synchronize the clocks of the \textit{nRF52832} over \gls{ble}. We could use highly stable crystals %the same crystals on both central and peripherals with the same load capacitors
% and synchronize these based on a \gls{ble} advertising packet, but this would result in a synchronization error increasing over time. We opted to actively synchronize these clocks.
%
%
% The first and major problem why we would need active time synchronization is the clock drift. Due to the inaccuracy of the crystal, temperature variations and frequency mismatch between central and peripherals, clock drift can occur. 
%
%
% \subsection{Time Independent Errors}
%
% Network time errors:~\cite{multi-unit-wireless-platform, rbs}
The total error rate on the synchronization is denoted as $\tau$. Several parameters contribute to this time error, depicted in Fig.~\ref{fig:causes_of_error-scematic}. For an accurate synchronization, $\tau$ has to be minimized.
The send time, i.e., the time to construct the message at the sending node, is deterministic.
The access time, i.e., the time between sending the message and the message actually being transmitted, can vary significantly in \gls{ble} networks. In a connection-based event with reliable data transfer, this varies from \SI{7.5}{ms} to \SI{4}{s}, while with advertising packets it can be in the order of microseconds. %A drawback of the latter is the reduced reliability of data transmission.
% We overcome this problem by proactively requesting time slots where the synchronization packet can be transmitted and only sample the reference time, by hardware implementation, at the beginning of each time slot. This almost completely eliminates the access time out of the equation.
The transmission time, i.e., the time to send a packet from sender to receiver, is dependent on the data rate, the packet length, and the channel conditions.
% The parameters above are important factors to minimize synchronization error at the master.
The wireless propagation time, e.g., \SI{30}{ns} for a distance of \SI{10}{m}, can be neglected.
% for a signal traveling from sender to receiver at the speed of light over a distance of \SI{10}{m} would be \SI{33}{ns}, which is negligible compared to the synchronization accuracy we need.
The reception time, i.e., the time needed for the peripheral to receive, decode and process the packet, is dependent on the used \gls{soc}.
% This is important at the receiver side.

% \Gls{ble} can operate differently depending on the application requirements. For reliable data transfer, \gls{ble} offers a connection-based mode where packets will be retransmitted in specific time slots until acknowledged. 
% % A disadvantage here is the long access time.
% For synchronization purposes, \gls{ble}'s advertising-based data transfer, with more deterministic latencies, yet less reliable data transmission, is used.
% % , which is commonly used to identify \gls{ble} devices, is used. 
% % This method features beacon based, fast communication with deterministic latencies, i.e. ideal for synchronization purposes, but suffers from unreliable data transfer. In ultra-low-power nodes, this communication method is commonly used.

\section{Synchronization Methodology}\label{sec:implementation_on_ble_soc}

% In \gls{ble} different communication methods can be used, connection based and advertising based (also called beacons). 
% We propose a mix of both methods to meet the expectations of our application. A high and reliable data transmission is needed for the \gls{imu} data. A disadvantage of this connection-based communication is that the data has to be transmitted in specific time slots, and has a long access time. We use beacon based, fast communication with deterministic latencies for synchronization purposes. 

% \jona{Introductie}

Following the analysis of causes of synchronization errors and identification of contributing components in the previous section, we here describe the methodology used to minimize these errors. % for a specific \gls{ble} implementation.
The \gls{dcu}'s main controller receives the accurate UTC time with second-level precision from the tablet.
The \gls{ble} chip on the \gls{dcu} replicates this time and establishes a relative synchronization between itself and the \glspl{imu} in a star network topology.
The relative synchronization is implemented using a simplified version of the \gls{ftsp} with MAC-layer time stamping, excluding clock drift compensation.
The master periodically sends synchronization packets to the slaves in a star network topology. Advertising packets are here used to decrease the access time as much as possible.
Since the master is not subject to any energy constraints, optimizing for energy consumption at this side is not necessary or relevant.
The slaves can optimize accuracy and energy consumption by intermittently listening for synchronization packets, and adjusting radio receive window lengths.
The resulting timestamps at the \gls{dcu} are calculated using a combination of the UTC time and relative \gls{imu} packet timestamps.
Since we do not compensate for clock drift, clocks are only updated during the transmission of a synchronization packet. Compared to \gls{ftsp}, our solution thus requires more transmissions, yet we achieve a sufficient accuracy for the application and incur only a very limited energy overhead at the \gls{imu} side. The accuracy is depicted in Fig.~\ref{fig:rtc_sync}. %\liesbet{depicts the synch error, not the power? would it be correct to state 'yet we achieve a sufficient accuracy for the application and incur only a very limited power overhead at the IMU side?}
%
% \sarah{Kan je dit wat meer verduidelijken?}\sarah{Wat is uiteindelijk het voordeel om deze methode te gebruiken ipv FTSP? Ook samenhangend met mijn bedenking later in de tekst, hoe vaak moet er bij FTSP opnieuw een synchronisatiepakket gestuurd worden? Ik had de indruk dat de moeilijkheid bij implementatie van FTSP zat in het kunnen nemen van timestamps op MAC layer level? Dit gebeurt hier ook, maar je past geen clock drift compensation toe en je krijgt een sub 1 µs nauwkeurigheid die dus eigenlijk nog net iets beter is dan FTSP? Waar zit dan precies het verschil met FTSP? Kan je dit één op één vergelijken zonder te vermelden wat het synchronisatie interval voor FTSP is? Dat is voor mij momenteel nog niet duidelijk, maar misschien begrijp ik het verkeerd? :p}\sarah{Kan je misschien wat meer benadrukken wat vernieuwend is aan deze methode ten opzichte van de state of the art?}
%
% A central node (\gls{dcu}) regularly transmits synchronization messages containing a timestamp. 
% We use a combination of \gls{mac} layer time stamping without clock drift compensation. 
Our solution does not interfere with any ongoing \gls{ble} communication, as it only requests available time slots in-between the normal \gls{ble} communication, i.e., when the channel is free.

\subsection{Timestamp Generation on Master}

The received UTC time with second-level accuracy from the tablet is transferred from the microcontroller (\textit{STM32}), in control of tablet communication, to the \gls{ble} radio (\textit{nRF}) over UART. The slight delay introduced here is not detrimental, since the synchronization between \glspl{imu} is handled by the \gls{ble} chips. %From this point on, the \gls{ble} radio uses a relative time synchronization and calculates the UTC time based on the received start time and the \gls{imu} packet timestamps.
The central node (\gls{ble} radio at the master) has a free-running timer. %All peripherals will synchronize their local timer to the master timer. %As discussed earlier, important parameters that influence synchronization accuracy at the master are send time, access time and transmission time.
For timekeeping, a combination of two \SI{16}{MHz} timers is used, derived from the \SI{32}{MHz} high-frequency crystal oscillator
%\footnote{muRata XRCGB32M000F1H19R0} 
with an accuracy of $\pm$\SI{10}{ppm}.
The wrap around value of 16 000, in combination with the \SI{62.5}{ns} resolution, corresponds to a wrap-around time of \SI{1}{ms}, which is used to accurately trigger transmission of \gls{imu} samples at different frequencies.
It takes \SI{49.71}{days} until a timer overflow occurs, i.e., plenty for typical \gls{wban} purposes.
%
% The first 32-bit timer is used as a timer with no prescaler (\SI{62.5}{ns} resolution) and a wrap-around value of 16 000. 
% The second 32-bit timer is used as a counter, counting the number of wraparounds of the first timer, hereby keeping track of time. 
% The structure of the \gls{ble} packet is shown in Fig.~\ref{fig:sync_packet}.
%
% \jona{Dit kan mss naar future work}
For a slightly lower power but less accurate time synchronization (each tick takes $\pm$~\SI{30}{\micro\second}), a clock derived from the \SI{32.768}{kHz} crystal, consuming only \SI{0.25}{\micro A}, instead of \SI{250}{\micro A}~\cite{nrf52832-datasheet} could be used. 
We chose to use the higher accuracy crystal, since this rather small increase in power consumption is negligible compared to the continuous \gls{ble} transmission.
%
% \begin{figure}[hbt!]
%     % \centering
%     \resizebox{0.9\columnwidth}{!}{%
%     % \centering
%         \input{sync_packet.tex}
%     }
%     \caption{Structure of the synchronization packet: Timer 1 is used as a timer with a resolution of \SI{62.5}{ns} with a \SI{1}{ms} wrap-around time, timer 2 is used in counter mode to count overflows of timer 1.}
%     \label{fig:sync_packet}
% \end{figure}
% \sarah{Ik ben niet thuis in de Nordic library, maar kan je verduidelijken wat er dan nieuw is aan deze methode? Ga je op een andere manier gebruik maken van bepaalde features binnen Nordic om hiermee ook te synchroniseren? Begrijp ik dan ook goed dat deze synchronisatie methode dan eigenlijk enkel gebruikt kan worden in combinatie met een Nordic chip? Zo ja, dan moet dit misschien ook reeds vermeld worden in het abstract?}
A Nordic library~\cite{nrf52_sync_nordic} was used, which implemented already a lot of features, such as the use of \gls{ppi} to set up radio transmission of sync packets, high-accuracy timestamping, and clock corrections. 
% A highly accurate MAC layer timestamp from the master increases the synchronization accuracy a lot compared to other approaches~\cite{ble_sync_connection_based_event}.
% Since we are using \gls{ble} connection-based \gls{imu} data transmission, the radio is not always free for transmitting synchronization packets. 
The Nordic Timeslot API allows the library to request time slots in between the normal \gls{ble} communication in advance, and herein transmit synchronization packets, i.e., minimizing the fixed access time to a couple of microseconds. 
The library was adapted to lower the energy consumption by carefully managing the high frequency clock states, radio receive window lengths, and update rates. At the peripheral, the interval between sync updates is managed by a timer, only enabling the receiver sporadically and disabling it after receiving a sync packet. 
The radio settings used are \SI{2}{MBit} PHY, $+$\SI{4}{dBm} output power.
The send and transmission time is kept predictable by capturing the timer value at a consistent time delta from the actual transmission. The fixed delay hereby introduced can be easily compensated for at the receiver. At the start of an allocated time slot, i.e., when the radio is ready, the timers will be captured by means of \gls{ppi}. The \gls{ppi} channels are synchronized to the \SI{16}{MHz} clock, meaning 1 clock cycle delay will be introduced here~\cite{nrf52832-datasheet}. When the radio is ready, the \gls{cpu} copies the \gls{ppi} sampled values from the timer registers into the transmit packet. The time this operation takes is predictable, depending on the compiler optimization level, and is in our case determined to be \SI{29}{\micro \second}. %This operation, depending on the compiler optimization level, will always take a predictable amount of time, i.e., \SI{29}{\micro \second}. % \liesbet{we assume or we know or we could determine the predicatble amount...?}
%Hereby, we can achieve a consistent time delta between sampling the timer values and the actual radio transmission.

\subsection{Timestamp Decoding at Peripheral}

% A second time-critical part is the reception time. 
It's important that the peripheral timers can be sampled as close to the packet reception as possible, while the actual time correction is not as time-critical, and can be postponed. 
Using the implementation of \cite{nrf52_sync_nordic}, the clocks are adjusted. A \gls{ppi} is set up to trigger the capture of both timers once the correct address is wirelessly received, generating a very accurate MAC-layer timestamp.
In the radio interrupt handler, the synchronization timer offset is calculated and compensated. A fixed delay is added to account for the introduced error at the master when copying the timer value buffers.
% The timer offset is calculated based on the local and master timestamps. This offset is stored in the timer compare register and correction of this timer value is also done using \gls{ppi}. 
Timer offset compensation is done using \gls{ppi} together with timer compare registers.
If the peripheral timer is ahead of the master, the current timer cycle is cut short and the counter timer is incremented. If the peripheral timer is behind the master timer, the current timer cycle is cut short without incrementing the counter, thus delaying the next \SI{1}{ms} pulse.

\section{Results}\label{sec:results}

% \subsection{Accuracy of the Proposed Method}

% \subsection{Accuracy and Energy Consumption}
% \sarah{Er is maar één subsectie in deze sectie?}

\begin{figure}[tbp]
    \centering
    \resizebox{0.8\columnwidth}{!}{%
        \definecolor{blue}{HTML}{9ac7f6}
\definecolor{orange}{HTML}{ffb179}
\definecolor{green}{HTML}{60eea3}
\definecolor{jona_red}{HTML}{ff9695}
\definecolor{purple}{HTML}{d9b2ff}
\definecolor{brown}{HTML}{e4bb98}
\definecolor{pink}{HTML}{ffa4e3}
\definecolor{grey}{HTML}{cfcfcf}
\definecolor{yellow}{HTML}{fbff9f}
\definecolor{aqua}{HTML}{a7f5f2}

\usetikzlibrary{shapes.misc}

% This file was created with tikzplotlib v0.9.17.
\begin{tikzpicture}[
legendnode/.style={cross out, draw=red, fill=white},
trim axis left, trim axis right
]

% \definecolor{blue}{rgb}{0.12156862745098,0.466666666666667,0.705882352941177}
\definecolor{color1}{rgb}{1,0.498039215686275,0.0549019607843137}

\begin{axis}[
use as bounding box,
legend cell align={left},
legend style={
  fill opacity=0.8,
  draw opacity=1,
  text opacity=1,
  at={(0.03,0.97)},
  anchor=north west,
  draw=white!80!black
},
log basis y={10},
tick align=outside,
tick pos=left,
title={RTC synchronization error},
x grid style={white!69.0196078431373!black},
xlabel={Time between receive windows [s] + IMU error},
xmajorgrids,
xmin=0.5, xmax=8.5,
xtick style={color=black},
xtick={1,2,3,4,5,6,7,8},
xticklabels={0,1,5,10,20,60,300,IMU},
y grid style={white!69.0196078431373!black},
ylabel={Synchronization error [us]},
ymajorgrids,
ymin=0.0986298447806941, ymax=10150.6502991114,
ymode=log,
ytick style={color=black},
ytick={0.001,0.01,0.1,1,10,100,1000,10000,100000,1000000},
yticklabels={
  \(\displaystyle {10^{-3}}\),
  \(\displaystyle {10^{-2}}\),
  \(\displaystyle {10^{-1}}\),
  \(\displaystyle {10^{0}}\),
  \(\displaystyle {10^{1}}\),
  \(\displaystyle {10^{2}}\),
  \(\displaystyle {10^{3}}\),
  \(\displaystyle {10^{4}}\),
  \(\displaystyle {10^{5}}\),
  \(\displaystyle {10^{6}}\)
}
]
\path [draw=black, fill=green]
(axis cs:0.75,0.499999998737621)
--(axis cs:1.25,0.499999998737621)
--(axis cs:1.25,1.33333332996699)
--(axis cs:0.75,1.33333332996699)
--(axis cs:0.75,0.499999998737621)
--cycle;
\addplot [black, forget plot]
table {%
1 0.499999998737621
1 0.01
};
\addplot [black, forget plot]
table {%
1 1.33333332996699
1 2.58333332681104
};
\addplot [black, forget plot]
table {%
0.875 0.1
1.125 0.1
};
\addplot [black, forget plot] % top of first plot
table {%
0.875 2.58333332681104
1.125 2.58333332681104
};

\path [draw=black, fill=blue]
(axis cs:1.75,2.29166664666991)
--(axis cs:2.25,2.29166664666991)
--(axis cs:2.25,6.54166672120482)
--(axis cs:1.75,6.54166672120482)
--(axis cs:1.75,2.29166664666991)
--cycle;
\addplot [black, forget plot]
table {%
2 2.29166664666991
2 0.1
};
\addplot [black, forget plot]
table {%
2 6.54166672120482
2 10.2083333359815
};
\addplot [black, forget plot]
table {%
1.875 0.1
2.125 0.1
};
\addplot [black, forget plot]
table {%
1.875 10.2083333359815
2.125 10.2083333359815
};
\path [draw=black, fill=orange]
(axis cs:2.75,17.9166666782749)
--(axis cs:3.25,17.9166666782749)
--(axis cs:3.25,53.4166666170677)
--(axis cs:2.75,53.4166666170677)
--(axis cs:2.75,17.9166666782749)
--cycle;
\addplot [black, forget plot]
table {%
3 17.9166666782749
3 0.1
};
\addplot [black, forget plot]
table {%
3 53.4166666170677
3 72.7916666667383
};
\addplot [black, forget plot]
table {%
2.875 0.1
3.125 0.1
};
\addplot [black, forget plot]
table {%
2.875 72.7916666667383
3.125 72.7916666667383
};
\path [draw=black, fill=jona_red]
(axis cs:3.75,34.7499999975298)
--(axis cs:4.25,34.7499999975298)
--(axis cs:4.25,103.999999964799)
--(axis cs:3.75,103.999999964799)
--(axis cs:3.75,34.7499999975298)
--cycle;
\addplot [black, forget plot]
table {%
4 34.7499999975298
4 0.1
};
\addplot [black, forget plot]
table {%
4 103.999999964799
4 139.291666666708
};
\addplot [black, forget plot]
table {%
3.875 0.1
4.125 0.1
};
\addplot [black, forget plot]
table {%
3.875 139.291666666708
4.125 139.291666666708
};
\path [draw=black, fill=purple]
(axis cs:4.75,41.5833332567672)
--(axis cs:5.25,41.5833332567672)
--(axis cs:5.25,124.166666637393)
--(axis cs:4.75,124.166666637393)
--(axis cs:4.75,41.5833332567672)
--cycle;
\addplot [black, forget plot]
table {%
5 41.5833332567672
5 0.1
};
\addplot [black, forget plot]
table {%
5 124.166666637393
5 168.166666666636
};
\addplot [black, forget plot]
table {%
4.875 0.1
5.125 0.1
};
\addplot [black, forget plot]
table {%
4.875 168.166666666636
5.125 168.166666666636
};
\path [draw=black, fill=brown]
(axis cs:5.75,104.50000002038)
--(axis cs:6.25,104.50000002038)
--(axis cs:6.25,325.833333313597)
--(axis cs:5.75,325.833333313597)
--(axis cs:5.75,104.50000002038)
--cycle;
\addplot [black, forget plot]
table {%
6 104.50000002038
6 0.166666609402455
};
\addplot [black, forget plot]
table {%
6 325.833333313597
6 427.20833332055
};
\addplot [black, forget plot]
table {%
5.875 0.166666609402455
6.125 0.166666609402455
};
\addplot [black, forget plot]
table {%
5.875 427.20833332055
6.125 427.20833332055
};
\path [draw=black, fill=yellow]
(axis cs:6.75,964.270833321734)
--(axis cs:7.25,964.270833321734)
--(axis cs:7.25,3111.31249998198)
--(axis cs:6.75,3111.31249998198)
--(axis cs:6.75,964.270833321734)
--cycle;
\addplot [black, forget plot]
table {%
7 964.270833321734
7 0.291666680141134
};
\addplot [black, forget plot]
table {%
7 3111.31249998198
7 5000.04166667622
};
\addplot [black, forget plot]
table {%
6.875 0.291666680141134
7.125 0.291666680141134
};
\addplot [black, forget plot]
table {%
6.875 5000.04166667622
7.125 5000.04166667622
};
\addplot [color1, forget plot]
table {%
0.75 0.833333331229369
1.25 0.833333331229369
};
\addplot [color1, forget plot]
table {%
1.75 4.41666668393736
2.25 4.41666668393736
};
\addplot [color1, forget plot]
table {%
2.75 35.6666666689875
3.25 35.6666666689875
};
\addplot [color1, forget plot]
table {%
3.75 69.5833333566043
4.25 69.5833333566043
};
\addplot [color1, forget plot]
table {%
4.75 82.9583333370465
5.25 82.9583333370465
};
\addplot [color1, forget plot]
table {%
5.75 215.874999923926
6.25 215.874999923926
};
\addplot [color1, forget plot]
table {%
6.75 2151.04166665014
7.25 2151.04166665014
};
\addplot [draw=red, fill=red, forget plot, mark=x, only marks]
table{%
x  y
1 0.66743827160493827160493827160494
2 20.023148148148148148148148148148
3 100.11574074074074074074074074074
4 200.23148148148148148148148148148
5 400.46296296296296296296296296296
6 1201.3888888888888888888888888889
7 6006.9444444444444444444444444444
};

\path [draw=black, fill=grey]
(axis cs:7.75,1097.14583335574)
--(axis cs:8.25,1097.14583335574)
--(axis cs:8.25,3288.07291668554)
--(axis cs:7.75,3288.07291668554)
--(axis cs:7.75,1097.14583335574)
--cycle;
\addplot [black, forget plot]
table {%
8 1097.14583335574
8 0.1
};
\addplot [black, forget plot]
table {%
8 3288.07291668554
8 4433.20833335292
};
\addplot [black, forget plot]
table {%
7.875 0.1
8.125 0.1
};
\addplot [black, forget plot]
table {%
7.875 4433.20833335292
8.125 4433.20833335292
};

\addplot [color1, forget plot]
table {%
7.75 2191.35416662652
8.25 2191.35416662652
};

\end{axis}

\node [legendnode,label=right:Theoretical max drift, fill=white] at (0.4,5.3) {};

\end{tikzpicture}
    % \includestandalone[width=\columnwidth]{rtc_sync}
    }
    \caption{\Gls{rtc} synchronization accuracy between 2 \gls{ble} nodes with different windows for receiving synchronization packets and \gls{imu} oversampling error. The 0-second receive window represents continuous reception.}
    \label{fig:rtc_sync}
    \vspace{-12pt}
\end{figure}
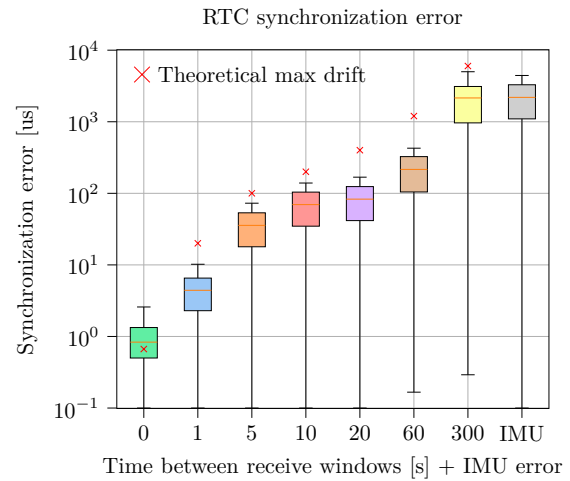

% A \SI{32}{MHz} \SI{6}{pF} quartz crystals\footnote{muRata XRCGB32M000F1H19R0} with a tolerance of \SI{10}{ppm} is used.
Fig.~\ref{fig:rtc_sync} shows the measured synchronization error and theoretical max derivation in clock stability, based on datasheet values. The synchronization error is measured over a period of \SI{30}{minutes} with a logic analyzer at \SI{24}{MSa/s} by toggling a \gls{gpio} pin by \gls{ppi} every \SI{10}{ms} from the synchronized clock at peripheral and master. Table~\ref{tab:power} depicts the measured additional energy consumption for synchronization.
% Every \SI{10}{ms}, the synchronization error is observed.
%
% A tradeoff can be made between energy consumption and synchronization accuracy.
%
The induced random error from oversampling the \gls{imu} is one of the biggest contributors to the total error.
When sending sync packets very frequently, i.e., at a fixed rate of \SI{30}{Hz}, it allows the receiver RX-window to be as short as possible. The lowest achievable synchronization error is less than \SI{1}{\micro\second}. This adds \SI{74.8}{\joule\per\hour} to the peripherals' energy consumption. Note that the theoretical maximal error is greater than the measured error due to lost synchronization packets when the wireless interface is busy transmitting \gls{ble} data packets.
It is worth noting that temperature variations can impact the frequency deviation of crystals, potentially deteriorating the accuracy.
The reception and processing of a single synchronization packet consumes \SI{3.34}{\milli\joule}.
When duty-cycling the receiver, i.e., enabling the receiver only until a sync packet is received, the synchronization error increases but energy consumption decreases. %A tradeoff can be made here.
%Using a \SI{10}{\second} interval, we obtain a synchronization error of sub \SI{100}{\micro\second} with a power consumption of \SI{334}{\micro\watt}. When increasing the interval to \SI{1}{\minute} an accuracy of $\pm$~\SI{200}{\micro\second} with a power consumption of \SI{55}{\micro\watt} is achieved.
Using a \SI{1}{\minute} interval, an accuracy of $\pm$~\SI{200}{\micro\second} with a energy consumption of \SI{198}{\milli\joule\per\hour} per hour is achieved.
Together with the maximal oversampling error, i.e., \SI{4.4}{ms}, we achieve a solution with an error of less than the predetermined \SI{5}{ms} maximal error.
% Worst case, when only receiving synchronization packets every \SI{5}{min}, the synchronization error of the clocks is 
% % comparable to the mean synchronization mismatch by oversampling the \gls{imu}, i.e., 
% equal to several milliseconds with a further reduced dedicated power consumption of \SI{10}{\micro\watt}. %In \gls{ftsp}, clock drift estimation ensures that far fewer synchronization packets are needed, further reducing energy consumption.
%
% When continuously enabling the receiver, the error is greater than the theoretical maximal clock drift due to lost synchronization packet when the wireless interface being busy transmitting \gls{ble} data packets.
%
% \subsection{Expansion to Synchronized \gls{emg}}
The synchronization architecture is designed in a way that new sensors, such as an \gls{emg} sensor can be easily added and synchronized with the same shared clock. The necessary \SI{2}{kHz} sampling rate can be derived from the synchronous clock, and can be used as a sampling trigger for the \gls{emg}. More synchronization packets can be transmitted to achieve sub-packet accuracy.

\begin{table}[htb!]
\vspace{-10pt}
\centering
\caption{Additional energy consumption for different rx windows per hour of operation}% in \si{\micro\watt} and \si{\second}}
\label{tab:power}
% \resizebox{\textwidth}{!}{%
    \begin{tabular}{@{}rccccccc@{}}
    \toprule
    Time [\si{\second}] & 0     & 1     & 5     & 10    & 20    & 60    & 300   \\ \midrule
    % Power [\si{\micro\watt}] & 20 790 & 3 338 & 667 & 334 & 166 & 55 & 11 \\
    Energy [\si{\milli\joule}] & 74 844 & 12 016 & 2 401 & 1 202 & 597 & 198 & 40 \\ \bottomrule
    \end{tabular}%
    \vspace{-12pt}
\end{table}

\section{Conclusion and Future Work}\label{sec:conclusion}

We presented a wirelessly synchronized multi-unit \gls{imu} platform capable of achieving sub \SI{1}{\micro\second} time accuracy between nodes. 
Further, an energy-optimized implementation with an accuracy of \SI{200}{\micro\second} and an energy consumption of only \SI{198}{\milli\joule\per\hour} is achieved.
% \liesbet{confusing accuracy statements, clarified with additions 'capable of achieving' above and 'power-optimized' below?}
% The proposed power-optimized implementation for synchronizing \gls{imu} packets wirelessly over \gls{ble} has an accuracy of \SI{200}{\micro\second} and only consumes \SI{55}{\micro\watt} of extra power.
Sources of synchronization error in an \gls{imu} system are discussed and minimized.
This approach suits the accuracy and low-power requirements for a multi-IMU system. 
In future work, the synchronization could be improved, both in terms of accuracy and energy consumption, by actively compensating for the clock drift and using a synchronized wake-up. %Both approaches could further reduce energy consumption at the peripheral and master.

% \jona{Dit kan mss naar future work}
% For a slightly lower-power but less accurate time synchronization, one could use a clock derived from the \SI{32.768}{kHz} crystal, consuming only \SI{0.25}{\micro A}, instead of \SI{250}{\micro A}~\cite{nrf52832-datasheet}. We chose to use the higher accuracy crystal, since this rather small increase in power consumption is negligible compared to the continuous \gls{ble} transmission, and a more accurate synchronization can be achieved.

% % \begin{thebibliography}{1}
% {\footnotesize%
% \bibliographystyle{IEEEtran}
% \bibliography{mybib} % Entries are in the refs.bib file
% }

{\footnotesize \printbibliography}

% \newpage

 

% \end{IEEEbiography}



%%%%%%%%%%%%%%%%%%%%%%%%%%%%%%%%%%%%%%%%%%%%%%%
% Publish in: IEEE Embedded Systems Letters
% https://ieee-ceda.org/publication/esl-publication/author-guidelines
%%%%%%%%%%%%%%%%%%%%%%%%%%%%%%%%%%%%%%%%%%%%%%%

\vfill

\end{document}